\documentclass[prc,aps,nofootinbib,showkeys,showpacs,preprint]{revtex4}
\usepackage{graphicx}  
\usepackage{epstopdf} 
\usepackage{dcolumn}   
\usepackage{bm}        
\usepackage{amssymb}   
\usepackage{amsmath}
\usepackage{longtable}

\begin{document}


\title{Analysis of correlations between dipole transitions $1^-_1\rightarrow 0^+_1$ and $3^-_1\rightarrow 2^+_1$ based on the collective model}

\author{R.V. Jolos$^{1,2,3}$ and E.A. Kolganova$^{1,2}$ }
\affiliation{$^1$Joint Institute for Nuclear Research, 141980 Dubna, Moscow region, Russia\\
$^2$Dubna State University, 141982 Dubna, Moscow Region, Russia\\
$^3$Lomonosov Moscow State University, Dubna branch, Russia}
\date{\today}

\begin{abstract}
\begin{description}
\item[Background:] Observed systematic deviations of the value of the $B(E1;1^-_1\rightarrow 0^+_1)/B(E1;3^-_1\rightarrow 2^+_1)$
ratio in spherical nuclei from the prediction of the collective model which take into account only quadrupole and octupole collective modes.
\item[Purpose:] Evaluate effect of the isovector dipole and quadrupole-octupole modes coupling on the
$B(E1;1^-_1\rightarrow 0^+_1)/B(E1;3^-_1\rightarrow 2^+_1)$ ratio.
\item[Methods:] The Hamiltonian of the phenomenological collective model is used to calculate mixing of the isovector dipole and quadrupole and octupole modes.
\item[Results:]  The effect of the admixture of the giant dipole resonance to the low-lying collective quadrupole and octupole modes is estimated.
\item[Conclusion: ] It is shown that the coupling of the quadrupole and octupole collective modes to giant dipole resonance leads to decrease of the ratio $B(E1;1^-_1\rightarrow 0^+_1)/B(E1;3^-_1\rightarrow 2^+_1)$ relative to the value 7/3 predicted by the pure collective quadrupole-octupole model.
\end{description}
\end{abstract}

\pacs{21.10.Re, 21.10.Ky, 21.60.Ev}
\maketitle

\section{Introduction}

In near magic nuclei and nuclei close to them in the number of valence nucleons, the lowest $1^-_1$ states have quadrupole-octupole two-phonon structure.
They were observed in many nuclei of this type \cite{Cottle,Zilges,Herzberg,Kneissl,Ibbotson,Fransen,Pietralla,Zilges1,Shneidman}.
The conclusion about their collective nature  was made on the number of facts \cite{Pietralla}  including excitation energies and electromagnetic transition probabilities,
especially, on the data on $B(E2;1^-_1 \rightarrow 3^-_1)/B(E2;2^+_1 \rightarrow0^+_1)$,  which show a smooth dependence on mass number. This property is characteristic of the collective states.

In deformed nuclei octupole excitations take place in an axially deformed nuclear mean field. As a result, four negative parity states with angular momentum projection 
on the symmetry axis K=0,1,2 and 3 appear in even-even nuclei. Rotational bands are based on these states,
and those of them with K=0 or 1 begin with $1^-$ states. A special group of nuclei is formed by isotopes of Rn, Ra, Th, U, and Pu in which $1^-$ states with unusually small excitation energies are observed.

The impressive correlation between $B(E1;1^-_1\rightarrow 0^+_1)$ and $B(E1;3^-_1\rightarrow 2^+_1)$ values have been reported
in \cite{Pietralla} basing on the analysis of the experimental data. Although the measured B(E1) values themselves vary within two orders of magnitude for different medium and heavy nuclei, the ratio $R\equiv B(E1;1^-_1\rightarrow 0^+_1)/B(E1;3^-_1\rightarrow 2^+_1)$ 
remains approximately constant with  a value near unity.
In was also shown in \cite{Pietralla} that in a simple bosonic model applied to description of $2^+_1, 3^-_1$, and $1^-_1$ states R is equal to 7/3. Later this result was reproduced in \cite{Jolos1,Jolos2} basing on the fermionic Q-phonon representation of the collective states \cite{Pietralla2,Pietralla3,Jolos3}. Let us also mention that an interesting result concerning the possible values of the ratio R has been obtained in \cite{Ponomarev} based on the quasiparticle Random Phase Approximation.

In this work, we analyze the ratio of the considered E1 transitions based mainly on the collective model although some elements of the microscopic approach are used also.

\section{Model}

In order to analyze the correlations between the probabilities of the E1 transitions  connecting low-lying collective states, we use a model that allows describing the properties of the $2^+_1, 3^-_1$, and $1^-_1$ states taking into account the contribution of the giant dipole resonance to their structure \cite{Vdovin}. The Hamiltonian of the model consists of three terms:
\begin{eqnarray}
\label{Eq1}
H=H_{quad}+H_{oct}+H_{dip}.
\end{eqnarray}
The first and the second terms describe quadrupole and octupole excitations in harmonic approximation:
\begin{eqnarray}
\label{Eq2}
H_{quad}=\omega_2\sum_{\mu}d^+_{2\mu}d_{2\mu},\nonumber\\
H_{oct}=\omega_3\sum_{\mu}f^+_{3\mu}f_{3\mu},
\end{eqnarray}
where $d^+_{2\mu}(d_{2\mu})$ and $f^+_{3\mu}(f_{3\mu})$ are quadrupole and octupole phonon creation (annihilation) operators, $\omega_2$ and $\omega_3$ are the energies of the $2^+_1$ and $3^-_1$ states, respectively. The term $H_{dip}$ describes giant dipole resonance mode and its coupling to quadrupole and octupole phonons.

The dipole mode Hamiltonian consists of two terms:
\begin{eqnarray}
\label{Eq3}
H_{dip}=H^{(0)}_{dip}+\frac{1}{2}\kappa_1\sum_{\mu}D^+_{1\mu}D_{1\mu},
\end{eqnarray}
where $H^{(0)}_{dip}$ is the Hamiltonian of independent nucleons with single particle states belonging neighboring shells:
\begin{eqnarray}
\label{Eq4}
H^{(0)}_{dip}=\sum_p(E_p-\lambda)a^+_pa_p+\sum_h(\lambda-E_h)a^+_ha_h.
\end{eqnarray}
Here $a^+_p(a^+_h)$ are particle (hole) creation operators. Index $p$ denotes single particle states of the unoccupied shell. Index $h$ denotes single particle state belonging to the valence (or completely occupied) shell.  $E_p$ and $E_h$ are single particle energies and $\lambda$ is the Fermi energy. $\kappa_1$ is a dipole interaction constant. Operator $D_{1\mu}$ is a dipole transition operator. It contains two terms. The first one:
\begin{eqnarray}
\label{Eq5}
D^{(1)}_{1\mu}=\sum_{p,h}p_{ph}\left((a^+_pa^+_h)_{1\mu}+(-1)^{\mu}(a_pa_h)_{1-\mu}\right)
\end{eqnarray}
generates the giant dipole resonance (GDR) through the residual interaction
\begin{eqnarray}
\label{Eq6}
\frac{1}{2}\kappa_1\sum_{\mu}(D^{(1)})^+_{1\mu}D^{(1)}_{1\mu}.
\end{eqnarray}
Here $p_{ph}$ is the single particle matrix element of the dipole operator. 
The second term
\begin{eqnarray}
\label{Eq7}
D^{(2)}_{1\mu}=\frac{\sqrt{35}}{3}(1+\chi)C_{LD}AZ(\alpha_2\alpha_3)_{1\mu}
\end{eqnarray}
contains a contribution of the quadrupole and octupole modes into dipole transition operator \cite{Strutinsky,BM}.
In (\ref{Eq7}) $\chi=-0.7$, $C_{LD}$=0.0007 fm \cite{Strutinsky,BM} and $\alpha_{2\mu}$, $\alpha_{2\mu}$  are quadrupole and octupole collective variables related to phonon creation and annihilation operators:
\begin{eqnarray}
\label{Eq8}
\alpha_{2\mu}=
\sqrt{\frac{\hbar}{2B_2\omega_2}}(d^+_{2\mu}+(-1)^{\mu}d_{2-\mu}),	\nonumber\\
\alpha_{3\mu}=\sqrt{\frac{\hbar}{2B_3\omega_3}}(f^+_{3\mu}+(-1)^{\mu}f_{3-\mu}).
\end{eqnarray}

Combining (\ref{Eq1})--(\ref{Eq8}) and neglecting the quadrupole-octupole coupling, we obtain the following Hamiltonian:
\begin{eqnarray}
\label{Eq9}
H=\omega_2\sum_{\mu}d^+_{2\mu}d_{2\mu}+\omega_3\sum_{\mu}f^+_{3\mu}f_{3\mu}\nonumber\\
+\sum_p(E_p-\lambda)a^+_pa_p+\sum_h(\lambda-E_h)a^+_ha_h+\frac{1}{2}\kappa_1\sum_{\mu}(D^{(1)}_{1\mu})^+D^{(1)}_{1\mu}\nonumber\\
+\kappa_1\sum_{\mu}(D^{(1)}_{1\mu})^+D^{(2)}_{1\mu}.
\end{eqnarray}
Let us consider the fifth term in (\ref{Eq9}). The dipole operator $D^{(1)}_{1\mu}$ creates particle-hole excitation  of the E1 type acting on the ground state of the Hamiltonian of the independent particle model. In the case of the even-even nucleus the part of the Hamiltonian (\ref{Eq9}), which includes third, fourth and fifth terms,  can be diagonalized in the Tamm-Dankoff approximation introducing collective isovector E1 phonon which is, in fact, the giant dipole resonance (GDR) phonon. Application of the Tamm-Dankoff approximation instead of RPA is justified by a large energy of the corresponding nonperturbed particle-hole excitations where particles and holes belong to different shells.

In the Tamm-Dankoff approximation the collective isovector dipole phonon creation operator $p^+_{1\mu}$ has the following structure:
\begin{eqnarray}
\label{Eq10}
p^+_{1\mu}=\sum_{p,h}\psi_{ph}(a^+_pa^+_h)_{1\mu}
\end{eqnarray}
with normalization condition
\begin{eqnarray}
\label{Eq11}
1=\sum_{p,h}\psi^2_{ph}.
\end{eqnarray}

From the equation of motion for the phonon operator $p^+_{1\mu}$ we obtain the following secular equation for the energy of the giant dipole resonance $\omega_{GDR}$:
\begin{eqnarray}
\label{Eq12}
1=\kappa_1\sum_{p,h}\frac{p^2_{ph}}{\omega_{GDR}-(E_p-E_h)},
\end{eqnarray}
and for the amplitude $\psi_{ph}$ we obtain:
\begin{eqnarray}
\label{Eq13}
\psi_{ph}=\Gamma\frac{p_{ph}}{\omega_{GDR}-(E_p-E_h)},
\end{eqnarray}
where $\Gamma$ is determined by the normalization condition
\begin{eqnarray}
\label{Eq14}
1=\Gamma^2\sum_{p,h}\frac{p^2_{ph}}{(\omega_{GDR}-(E_p-E_h))^2}.
\end{eqnarray}

To simplify consideration, we make the following approximation: we neglect the energy
splitting of the single particle levels in each shell, putting $(E_p-E_h)$ equal to $\omega_0=41 A^{-1/3}$ MeV.
Then from (\ref{Eq12}) and (\ref{Eq14}) we obtain:
\begin{eqnarray}
\label{Eq15}
1=\kappa_1\frac{\sum_{p,h}p^2_{ph}}{\omega_{GDR}-\omega_0},
\end{eqnarray}
\begin{eqnarray}
\label{Eq16}
1=\Gamma^2\frac{\sum_{p,h}p^2_{ph}}{(\omega_{GDR}-\omega_0)^2}.
\end{eqnarray}
It follows from (\ref{Eq15}) and (\ref{Eq16}) that
\begin{eqnarray}
\label{Eq17}
\Gamma^2=\kappa_1(\omega_{GDR}-\omega_0).
\end{eqnarray}
The particle-hole creation operator presented in (\ref{Eq10}) can be expressed in terms of the collective and noncollective phonon creation operators:
\begin{eqnarray}
\label{Eq18}
(a^+_pa^+_h)_{1\mu}=\psi_{ph}p^+_{1\mu}+ ...
\end{eqnarray}
where ellipsis in (\ref{Eq18}) present noncollective $1^-$ states.
Substituting (\ref{Eq18}) into (\ref{Eq5}) and using (\ref{Eq12}) and (\ref{Eq13}) we obtain:
\begin{eqnarray}
\label{Eq19}
D^{(1)}_{1\mu}=\frac{\Gamma}{\kappa_1}(p^+_{1\mu}+(-1)^{\mu}p_{1-\mu}).
\end{eqnarray}

The sum of the third, fourth and fifth terms of the Hamiltonian (\ref{Eq9}) can be presented in the case of even-even nuclei as
\begin{eqnarray}
\label{Eq20}
\omega_{GDR}\sum_{\mu}p^+_{1\mu}p_{1\mu}.
\end{eqnarray}

Substituting (\ref{Eq19}), (\ref{Eq20}), (\ref{Eq7}) and (\ref{Eq8}) into (\ref{Eq9}) we obtain:
\begin{eqnarray}
\label{Eq21}
H=\omega_2\sum_{\mu}d^+_{2\mu}d_{2\mu}+\omega_3\sum_{\mu}f^+_{3\mu}f_{3\mu}+\omega_{GDR}\sum_{\mu}p^+_{1\mu}p_{1\mu}\nonumber\\
+\Gamma C\left(\sqrt{3}\sum_{\mu}(-1)^{\mu}p^+_{1\mu}(d^+_2f^+_3)_{1-\mu}+\sqrt{\frac{3}{7}}\sum_{\mu}(p^+_1d^+_2)_{3\mu}f_{3\mu}\right.\nonumber\\
\left.-\sqrt{\frac{3}{5}}\sum_{\mu}(p^+_1f^+_3)_{2\mu}d_{2\mu}+\sum_{\mu}p^+_{1\mu}(d_2f_3)_{1\mu}+h.c.\right),
\end{eqnarray}
where
\begin{eqnarray}
\label{Eq22}
C\equiv\frac{\sqrt{35}}{3}(1+\chi)C_{LD}AZ.
\end{eqnarray}

In the case where the residual interaction is not factorized, as suggested above, but has a more complicated form, its matrix elements will not necessarily be in phase with the matrix elements of the factorized dipole interaction considered above. This fact will weaken a mixing in the wave 
functions of the considered states of the dipole and the quadrupole-octupole components. This effect can be effectively accounted for by replacing the parameter $C$ in the Hamiltonian (\ref{Eq21}), which is included in the expression for the E1 transition operator, with the smaller in absolute value parameter $C'$  having the same sign as the parameter $C$.

Using the Hamiltonian (\ref{Eq21}) where we replace the parameter $C$ with $C'$ we find the vectors of the low-lying collective states of interest to us, namely, $2^+_1, 3^-_1, 1^-_1$,  and the ground state. Given the large excitation energy of the giant dipole resonance we can  limit calculation to the following approximate expressions for the desired wave vectors:
\begin{eqnarray}
\label{Eq23}
|1^-_1,M\rangle&=&\sqrt{1-x^2_1}(d^+_2f^+_3)_{1M}|0\rangle+x_1p^+_{1M}|0\rangle,\nonumber\\
|2^+_1,M\rangle&=&\sqrt{1-x^2_2}d^+_{2M}|0\rangle+x_2(p^+_1f^+_3){2M}|0\rangle,\nonumber\\
|3^-_1,M\rangle&=&\sqrt{1-x^2_3}f^+_{3M}|0\rangle+x_3(p^+_1d^+_2){3M}|0\rangle,\nonumber\\
|0^+_1\rangle&=&\sqrt{1-x^2_0}|0\rangle+x_0(p^+_1(d^+_2f^+_3)_1)_0|0\rangle.
\end{eqnarray}
The relationships between the interaction constant and the energy differences between the mixing states allow to use perturbation theory. In this approximation:
\begin{eqnarray}
\label{Eq24}
x_1&\approx&-\frac{\Gamma C'}{\omega_{GDR}-\omega_2-\omega_3},\nonumber\\
x_2&\approx& \sqrt{\frac{3}{5}}\frac{\Gamma C'}{\omega_{GDR}+\omega_3-\omega_2},\nonumber\\
x_3&\approx& -\sqrt{\frac{3}{7}}\frac{\Gamma C'}{\omega_{GDR}+\omega_2-\omega_3},\nonumber\\
x_0&\approx& -\sqrt{3}\frac{\Gamma C'}{\omega_{GDR}+\omega_2+\omega_3},
\end{eqnarray}
Combining (\ref{Eq5}), (\ref{Eq7}), (\ref{Eq8}), and (\ref{Eq19}) we obtain the following expression for the E1 transition operator:
\begin{eqnarray}
\label{Eq25}
D_{1\mu}=\frac{\Gamma}{\kappa_1}(p^+_{1\mu}+(-1)^{\mu}p_{1-\mu})
+C\left((d^+_2f^+_3)_{1\mu}+(-1)^{\mu}(d_2f_3)_{1-\mu}\right.\nonumber\\
\left.-\sqrt{\frac{3}{5}}\sum_{\nu,\lambda}C^{2\nu}_{3\lambda  1\mu}d^+_{2\nu}f_{3\lambda}+\sqrt{\frac{3}{7}}\sum_{\nu,\lambda}C^{3\nu}_{2\nu  1\mu}f^+_{3\lambda}d_{2\nu}\right).
\end{eqnarray}
In the expression (\ref{Eq25}) the factor, which will be reduced when calculating the  ratio of the E1 transition probabilities, is omitted.
\begin{table}
\caption{The results of calculations of the ratio $R\equiv B(E1;1^-_1\rightarrow 0^+_1)/B(E1;3^-_1\rightarrow 2^+_1)$. The calculations are performed for two values of the quantity $C'/C$: 1.0 and 0.933. The ratios in the final column were fitted to the experimental $^{142}$Nd value. For comparison are shown the experimental data, which are taken from \cite{Herzberg}, \cite{Ibbotson},\cite{Pietralla} and \cite{nndc}.}
\begin{tabular}{c|c|c|c|c|c}
\hline
Nucleus & $\omega_2$(MeV) &$\omega_3$(MeV) & $R_{exp}$ & $R_{cal}, \frac{C'}{C}$=1.0 & $R_{cal}, \frac{C'}{C}$=0.933 \\
\hline
$^{88}$Sr & 1.836 & 2.734 & 1.18 $\pm$ 0.26 & 0.34  & 0.90 \\
$^{106}$Pd& 0.512 & 2.084 & $2.10 \pm 0.95$   & 1.57  & 1.89 \\
$^{116}$Sn& 1.294 & 2.266 & $1.29 \pm 0.47$ & 0.46  & 1.14 \\
$^{118}$Sn& 1.230 & 2.325 & $1.04 \pm 0.19$ & 0.44  & 1.13 \\
$^{120}$Sn& 1.171 & 2.400 & $1.25 \pm 0.13$ & 0.41  & 1.12 \\
$^{122}$Sn& 1.141 & 2.493 & $1.09 \pm 0.11$ & 0.34  & 1.08 \\
$^{124}$Sn& 1.132 & 2.602 & $1.00 \pm 0.14$  & 0.25  & 1.00 \\
$^{140}$Ce& 1.596 & 2.464 & $0.41 \pm 0.01$ & 0.0065& 0.540\\
$^{142}$Nd& 1.576 & 2.084 & $0.77 \pm 0.04$ & 0.045 & 0.77 \\
$^{144}$Nd& 0.697 & 1.511 & $1.78 \pm 0.23$ & 1.21  & 1.75 \\
$^{144}$Sm& 1.660 & 1.810 & $1.12 \pm 0.26$ & 0.10  & 0.87 \\
$^{148}$Nd& 0.302 & 0.999 & $3.39 \pm 1.57$ & 1.97  & 2.15 \\                     
\hline
\end{tabular}
\end{table}
\normalsize

Using (\ref{Eq23}-\ref{Eq25}) we obtain the following result:
\begin{eqnarray}
\label{Eq26}
R\equiv\frac{B(E1;1^-_1\rightarrow 0^+_1)}{B(E1;3^-_1\rightarrow 2^+_1)}=\frac{7}{3}\left(\frac{1-\frac{C}{C'}\frac{2\omega_{GDR}(\omega_{GDR}-\omega_0)}{\omega^2_{GDR}-(\omega_2+\omega_3)^2}}{1-\frac{C}{C'}\frac{2\omega_{GDR}(\omega_{GDR}-\omega_0)}{\omega^2_{GDR}-(\omega_3-\omega_2)^2}}\right)^2
\end{eqnarray}
The results of calculation of $R$ are presented in Table 1 and Fig. 1 for two values of the ratio $C'/C =$ 1 and 0.933.
The last value is fixed to reproduce the experimental value of $R$ for $^{142}$Nd. In the calculation we use the following parametrization  $\omega_{GDR} = 94.68\left(A^{-1/3}-A^{-2/3}\right)$ MeV ~\cite{Firestone}.

As it is seen from the presented results an admixture of the giant dipole resonance to the wave functions of the low-lying collective states explain the decreasing of the ratio $R$ relative to the value 7/3 predicted by the collective model taking into account only quadrupole and octupole modes. It is seen also that this decrease of $R$ is smaller in nuclei with smaller values of the excitation energies of the $2^+_1$ and $3^-_1$ states.  
\begin{figure}[t]
\resizebox{0.95\textwidth}{!}{%
  \includegraphics{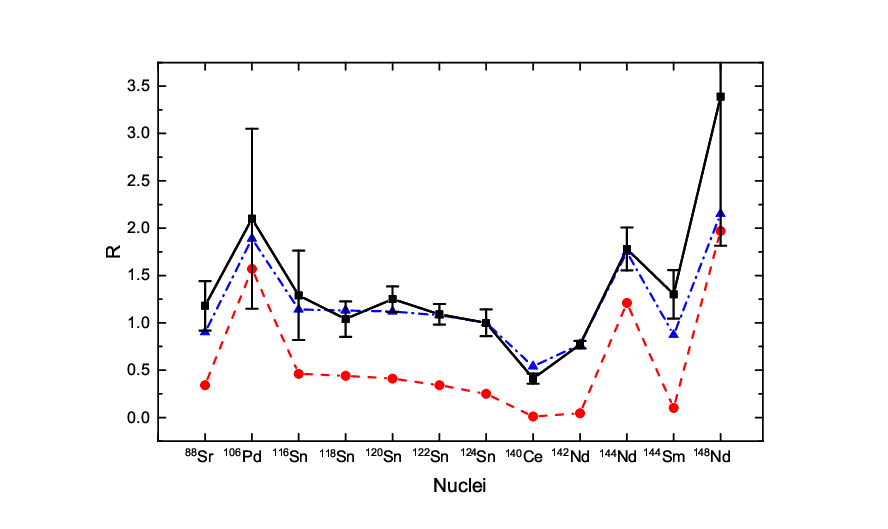}
}
\caption{The results of calculations of the ratio $R\equiv B(E1;1^-_1\rightarrow 0^+_1)/B(E1;3^-_1\rightarrow 2^+_1)$. The calculations are performed for two values of the quantity $C'/C$: 1.0 (red circle, dashed line) and 0.933 (blue triangle, dashed-dotted line). For comparison are shown the experimental data (black square, solid line), which are taken from \cite{Herzberg}, \cite{Ibbotson}, \cite{Pietralla} and \cite{nndc}.}
\label{fig:2}       
\end{figure}

\section{Conclusion}

The Hamiltonian of the phenomenological collective model has been used 
to calculate mixing of the isovector dipole and low-lying quadrupole-octupole modes. 
Taking into account the admixture of the giant dipole resonance to the wave functions of the low-lying  collective states we obtained a simple relation for the ratio of the E1 transition probabilities. 
It is shown that the coupling of the quadrupole and octupole collective modes to the giant dipole resonance leads to a decrease of the ratio $B(E1;1^-_1\rightarrow 0^+_1)/B(E1;3^-_1\rightarrow 2^+_1)$ relative to the value 7/3 predicted by the pure quadrupole-octupole collective model~\cite{Pietralla}.  The effect is larger for nuclei with larger excitation energies of the $2^+_1$ and $3^-_1$ states. In the calculations performed in this work all high lying $1^-$ states usually interpreted as the
giant dipole resonance are replaced by one collective state. Realistic calculations require to include into consideration all high lying $1^-$ states distributed over a wide energy interval.

\acknowledgments{
The authors are grateful to Prof. A.I. Vdovin for useful discussion.
}


\end{document}